\documentstyle[12pt]{article}
\begin{document}
\vspace{1.cm}
\begin{center}
\Large\bf Radiative Decays of Heavy Hadrons From Light-Cone\\
QCD Sum Rules in the Leading Order of HQET 
\end{center}
\vspace{0.5cm}
\begin{center}
{Shi-Lin Zhu and Yuan-Ben Dai}\\\vspace{3mm}
{\it Institute of Theoretical Physics, 
 Academia Sinica, P.O.Box 2735, Beijing 100080, China }
\end{center}

\vspace{1.cm}
\begin{abstract}
The radiative decays of heavy baryons and lowest three doublets of 
heavy mesons are studied with the light cone QCD sum rules in the 
leading order of heavy quark effective theory. 
\end{abstract}



\vspace{1.cm}

\pagenumbering{arabic}

\section{Introduction}
\label{sec1} 
Heavy quark effective theory (HQET) \cite{grinstein} provides a framework
to study the heavy hadron spectra and transition amplitude with the 
systematic expansion in terms of $1/ m_Q$. 
Using the light cone photon wave function (PWF), 
radiative decay processes like $B\to {\sl l} \nu \gamma$, $B\to \rho \gamma$
have been studied \cite{mendel,aliev,ali,stoll,ball} with QCD sum rules (QSR) \cite{svz}. 
Recently similar approach was employed to analyze
the couplings of pions with heavy hadrons \cite{zhu}. 
In this work we will study 
the radiative decays of heavy baryons and lowest three doublets of 
heavy mesons using the light cone QSR (LCQSR) \cite{braun-lc} in the 
leading order of HQET. With LCQSR the continuum and excited 
states contribution is subtracted more cleanly, which is in contrast 
with the analysis of meson radiative decays using the external field 
method in QSR \cite{zhu-rad}.
The LCQSRs for the radiative decays of heavy baryons are presented 
in section \ref{sec2}. Section \ref{sec3} discusses radiative decays
of lowest three doublets of heavy mesons. The following section 
is a discussion of the parameters and the photon wave functions.
The last section is the numerical analysis and a short summary. 

%%%%%%%%%%%%%%%%%%%%%%%%%%%%%%%%%%%%%%%%%%%%%%%%%%%%%%%%%%%%%%%%%%%%%%%%%%%%%%%%
\section{Radiative decays of heavy baryons}
\label{sec2}
We first introduce the interpolating currents for the heavy baryons:
\begin{equation}
\label{la1}
\eta_{\Lambda} (x) =\epsilon_{abc} [{u^a}^T(x) C\gamma_5 d^b (x)] h_v^c (x) \, ,
\end{equation}
\begin{equation}
\label{si1}
\eta_{\Sigma^+} (x) =\epsilon_{abc} [{u^a}^T(x) C\gamma_{\mu} d^b (x)] 
\gamma^{\mu}_t \gamma_5 h_v^c (x) \, ,
\end{equation}
\begin{equation}
\label{star1}
\eta^{\mu}_{{\Sigma^{++}}^*} (x) =\epsilon_{abc} [{u^a}^T(x) C\gamma_{\nu} u^b (x)] 
(-g_t^{\mu\nu}+ {1\over 3} \gamma_t^{\mu}\gamma_t^{\nu} )h_v^c (x) \, ,
\end{equation}
where $a$, $b$, $c$ is the color index, $u(x)$, $d(x)$, $h_v(x)$ is the up,
down and heavy quark fields, $T$ denotes the transpose, $C$ is the charge conjugate 
matrix,
$g_t^{\mu\nu}=g^{\mu\nu}-v^{\mu}v^{\nu}$, $\gamma_t^{\mu}=\gamma_{\mu}-
{\hat v}v^{\mu}$, and $v^{\mu}$ is the velocity of the heavy hadron.

The overlap amplititudes of the interpolating currents with the heavy baryons 
is defined as:
\begin{equation}
\label{la2}
\langle 0|\eta_{\Lambda} |\Lambda\rangle =f_{\Lambda}u_{\Lambda} \, ,
\end{equation}
\begin{equation}
\label{si2}
\langle 0|\eta_{\Sigma} |\Sigma\rangle =f_{\Sigma} u_{\Sigma}\, ,
\end{equation}
\begin{equation}
\label{star2}
\langle 0|\eta^{\mu}_{\Sigma^*} |\Sigma^*\rangle ={f_{\Sigma^*}\over \sqrt{3}}
 u^{\mu}_{\Sigma^*} \, ,				
\end{equation}
where $u^{\mu}_{\Sigma^*}$ is the Rarita-Schwinger spinor in HQET. 
In the leading order of HQET, $f_{\Sigma}= f_{\Sigma^*}$ \cite{mass}.

The coupling constants $\eta_i$ are defined through the following amplitudes:
\begin{equation}
\label{coup-1}
M(\Sigma_c\to \Lambda_c \gamma )=ie\eta_1 
{\bar u}_{\Lambda_c} \sigma^{\mu\nu} q_\mu e_\nu u_{\Sigma_c} \; ,
\end{equation}
\begin{equation}
\label{coup-2}
M(\Sigma^*_c\to \Lambda_c \gamma )=ie\eta_2 
\epsilon_{\mu\nu\alpha\beta}
{\bar u}_{\Lambda_c}\gamma^\nu q^\alpha e^\beta u^\mu_{\Sigma^*_c} \; ,
\end{equation}
\begin{equation}
\label{coup-3}
M(\Sigma^*_c\to \Sigma_c \gamma )=ie\eta_3 
\epsilon_{\mu\nu\alpha\beta}
{\bar u}_{\Sigma_c} \gamma^\nu q^\alpha e^\beta u^\mu_{\Sigma^*_c} \; ,
\end{equation}
where $e_\mu$ and $q_\mu$ are the photon polarization vector and momentum
respectively, $e$ is the charge unit.

In order to derive the sum rules for the coupling constants we consider 
the correlator 
\begin{equation} 
\label{lam1}
 \int d^4x\;e^{-ik\cdot x}\langle\gamma(q)|T\left(\eta_{\Sigma}(0)
 {\bar \eta}_{\Lambda}(x)\right)|0\rangle =e
   {1+{\hat v}\over 2} \gamma_t^\mu \gamma_5 
  \epsilon_{\mu\alpha\beta\sigma} e^\alpha q^\beta v^\sigma 
   G_{\Sigma ,\Lambda}(\omega,\omega') \;,
\end{equation}
\begin{eqnarray}\nonumber
\label{sig1}
 \int d^4x\;e^{-ik\cdot x}\langle\gamma(q)|T\left(\eta^{\mu}_{\Sigma^*}(0)
{\bar \eta}_{\Lambda}(x)\right)|0\rangle = & \\
e(e_\alpha q_\nu -e_\nu q_\alpha )
 {1+{\hat v}\over 2}(-g_t^{\mu\nu}+{1\over 3}\gamma_t^{\mu}\gamma_t^{\nu}) 
 \epsilon_{\nu\alpha\beta\sigma} e^\alpha q^\beta v^\sigma  
   G_{\Sigma^* ,\Lambda}(\omega,\omega') &\;,
\end{eqnarray}
\begin{eqnarray}\nonumber
\label{sig2}
 \int d^4x\;e^{-ik\cdot x}\langle\gamma(q)|T\left(\eta^{\mu}_{\Sigma^*}(0)
{\bar \eta}_{\Sigma}(x)\right)|0\rangle = & \\
e{1+{\hat v}\over 2} \gamma_t^{\alpha}\gamma_5 
 (-g_t^{\mu\nu}+{1\over 3}\gamma_t^{\mu}\gamma_t^{\nu}) 
(e_\alpha q_\nu -e_\nu q_\alpha )   G_{\Sigma^* ,\Sigma}(\omega,\omega') &\;,
\end{eqnarray}
where $k^{\prime}=k-q$, $q^t_\mu =q_\mu -(q\cdot v) v_\mu$, 
$\omega=2v\cdot k$, $\omega^{\prime}=2v\cdot k^{\prime}$ and $q^2=0$. 

Let us first consider the functions $G_{\Sigma ,\Lambda}(\omega,\omega^{\prime})$ etc 
in (\ref{lam1})-(\ref{sig2}). As functions of two variables, they have the following 
pole terms from double dispersion relation 
\begin{eqnarray}
\label{pole1}
{-4i\eta_1 f_{\Sigma}f_{\Lambda}\over (2\bar\Lambda_{\Sigma}
-\omega')(2\bar\Lambda_{\Lambda}-\omega)}+{c\over 2\bar\Lambda_{\Sigma}
-\omega'}+{c'\over 2\bar\Lambda_{\Lambda}-\omega}\;,
\end{eqnarray}
\begin{eqnarray}
\label{pole2}
{-4i\eta_2 \over \sqrt{3} }{f_{\Sigma^*}f_{\Lambda}\over (2\bar\Lambda_{\Sigma^*}
-\omega')(2\bar\Lambda_{\Lambda}-\omega)}+{c\over 2\bar\Lambda_{\Sigma^*}
-\omega'}+{c'\over 2\bar\Lambda_{\Lambda}-\omega}\;,
\end{eqnarray}
\begin{eqnarray}
\label{pole3}
{4\eta_3 \over \sqrt{3}}{f_{\Sigma^*}f_{\Sigma}\over (2\bar\Lambda_{\Sigma^*}
-\omega')(2\bar\Lambda_{\Sigma}-\omega)}+{c\over 2\bar\Lambda_{\Sigma^*}
-\omega'}+{c'\over 2\bar\Lambda_{\Sigma}-\omega}\;,
\end{eqnarray}
where $f_{\Sigma^*}$ etc are constants defined in 
(\ref{la2})-(\ref{star2}), $\bar\Lambda_{\Sigma^*}=m_{\Sigma^*}-m_Q$. 

Keeping the two particle component of the photon wave function, 	
the expression for $G_{\Sigma^* , \Lambda}(\omega, \omega')$ 
with the tensor structure reads
\begin{eqnarray}\label{lam5}\nonumber			
-2  \int_0^{\infty} dt \int dx e^{-ikx} 
\delta (-x-vt){\bf Tr} \{ 
[C \langle \gamma (q)|u(0) {\bar u}(x) |0\rangle^T C \gamma_\mu iS(-x) \gamma_5 ]
&\\
+[ C iS^T(-x) C\gamma_{\mu}] \langle \gamma (q)|u(0) {\bar u}(x) |0\rangle
\gamma_5 ]\} &\; ,
\end{eqnarray}
where $iS(-x)$ is the full light quark propagator with both perturbative  
term and contribution from vacuum fields. 
\begin{eqnarray}\label{prop}\nonumber
iS(x)=\langle 0 | T [q(x), {\bar q}(0)] |0\rangle 
=i{{\hat x}\over 2\pi^2 x^4} -{\langle {\bar q} q\rangle  \over 12}
-{x^2 \over 192}\langle {\bar q}g_s \sigma\cdot G q\rangle &\\ 
-ig_s{1\over 16\pi^2}\int^1_0 du \{
{{\hat x}\over x^2} \sigma\cdot G(ux)-4iu {x_\mu\over x^2} 
G^{\mu\nu}(ux)\gamma_\nu \} +\cdots  & \; .
\end{eqnarray}

The light cone two-particle photon wave functions are \cite{mendel}:
\begin{eqnarray}\label{wf-1}\nonumber
<\gamma (q)| {\bar q} (x) \sigma_{\mu\nu} q(0) |0>=
i e_q e \langle {\bar q} q\rangle 
\int_0^1 du  e^{iuqx}\{ (e_\mu q_\nu -e_\nu q_\mu )  [\chi \phi (u) +x^2 g_1(u)]
&\\
+[ (qx)(e_\mu x_\nu -e_\nu x_\mu ) +(ex)(x_\mu q_\nu -x_\nu q_\mu )
-x^2(e_\mu q_\nu -e_\nu q_\mu )]g_2(u)\}  &\; ,
\end{eqnarray} 
\begin{equation}\label{wf-2}
<\gamma (q)| {\bar q} (x) \gamma_\mu \gamma_5 q(0) |0>=
{f\over 4} e_q e \epsilon_{\mu\nu\rho\sigma}e^\nu q^\rho x^\sigma  
\int_0^1 du  e^{iuqx} \psi (u) \; .
\end{equation} 
Due to the choice of the gauge  $x^\mu A_\mu(x) =0$, the path-ordered gauge 
factor \\
$P \exp\big(i g_s \int_0^1 du x^\mu A_\mu(u x) \big)$ has been omitted. 
The $\phi (u), \psi (u)$ is associated with the leading twist two photon 
wave function, while $g_1(u)$ and $g_2(u)$ are twist-4 PWFs.
All these PWFs are normalized to unity, $\int_0^1 du \; f (u) =1$.

Expressing (\ref{lam5}) with the photon wave functions, we arrive at:
\begin{eqnarray}\label{quark1}\nonumber
G_{\Sigma^* , \Lambda}(\omega, \omega')=
-(e_u -e_d)  \int_0^{\infty} dt 
\int_0^1 du e^{i (1-u) {\omega t \over 2}} e^{i u {\omega' t \over 2}} 
\{
\langle {\bar q} q\rangle [{1\over \pi^2 t^3} \chi \phi (u) &\\ 
+{1\over \pi^2 t} (g_1(u) -g_2 (u))]
+{f\over 24} \psi (u) t (\langle {\bar q} q \rangle +{t^2\over 16} 
\langle {\bar q}g_s\sigma\cdot G q \rangle ) \} +\cdots &  \;.
\end{eqnarray}

Similarly we have,
\begin{equation}\label{quark2}
G_{\Sigma, \Lambda}(\omega, \omega')=G_{\Sigma^* , \Lambda}(\omega, \omega')\; ,
\end{equation}
\begin{eqnarray}\label{quark3}\nonumber
G_{\Sigma^* , \Sigma}(\omega, \omega')= (e_u +e_d)
\int_0^{\infty} dt  \int_0^1 du e^{i (1-u) {\omega t \over 2}}
e^{i u {\omega' t \over 2}} \{ {f\over 4\pi^2 t^2} \psi (u) 
&\\
+{\langle {\bar q} q \rangle \over 6}(\langle {\bar q} q \rangle +{t^2\over 16} 
\langle {\bar q}g_s\sigma\cdot G q \rangle )
[\chi \phi (u) +t^2 (g_1(u)-g_2(u)] \}+\cdots &  \;,
\end{eqnarray}
where $\langle {\bar q} q \rangle=-(225\mbox{MeV})^3$, 
$\langle {\bar q}g_s\sigma\cdot G q \rangle =m_0^2\langle {\bar q} q \rangle$, 
$m_0^2=0.8$GeV$^2$. 
For large Euclidean values of $\omega$ and $\omega'$ 
this integral is dominated by the region of small $t$, therefore it can be 
approximated by the first a few terms.

After Wick rotations and making double Borel transformation 
with the variables $\omega$ and $\omega'$
the single-pole terms in (\ref{pole1})-(\ref{pole3}) are eliminated. 
Subtracting the continuum contribution which is modeled by the 
dispersion integral in the region 
$\omega ,\omega' \ge \omega_c$, we arrive at:
\begin{eqnarray}\label{final-b-1}
\nonumber
 \eta_1 f_{\Sigma} f_{\Lambda} = &- {1\over 64\pi^4} (e_u -e_d)a 
 e^{ {\Lambda_{\Sigma} +\Lambda_{\Lambda}  \over T }}
 \{
\chi \phi (u_0) T^4 f_3({\omega_c\over T}) \\
&
-4 [g_1(u_0)-g_2(u_0)]T^2f_1({\omega_c\over T})
+{2\pi^2\over 3} f\psi (u_0) (1-{m_0^2\over 4T^2}) \} &\;,
\end{eqnarray}
\begin{eqnarray}\label{final-b-2}
\nonumber
 \eta_2 f_{\Sigma^*} f_{\Lambda} = &- {\sqrt{3}\over 64\pi^4}(e_u-e_d) a 
 e^{ {\Lambda_{\Sigma^*} +\Lambda_{\Lambda}  \over T }}
 \{
\chi \phi (u_0) T^4 f_3({\omega_c\over T}) \\
&
-4 [g_1(u_0)-g_2(u_0)]T^2f_1({\omega_c\over T})
+{2\pi^2\over 3}f \psi (u_0) (1-{m_0^2\over 4T^2}) \} &\;,
\end{eqnarray}
\begin{eqnarray}\label{final-b-3}
\nonumber
 \eta_3 f_{\Sigma^*} f_{\Sigma} = & {\sqrt{3}\over 32\pi^2}(e_u+e_d)  
 e^{ {\Lambda_{\Sigma^*} +\Lambda_{\Sigma}  \over T }}
 \{
f \psi (u_0) T^3 f_2({\omega_c\over T}) \\
&
-{a^2\over 6\pi^2} (1-{m_0^2\over 4T^2}) 
[\chi \phi (u_0)T f_0 ({\omega_c\over T}) -{4\over T}\left(g_1(u_0)-g_2(u_0)\right)]
\} &\;,
\end{eqnarray}
where $f_n(x)=1-e^{-x}\sum\limits_{k=0}^{n}{x^k\over k!}$ is the factor used 
to subtract the continuum, $\omega_c$ is the continuum threshold.
$u_0={T_1 \over T_1 + T_2}$, 
$T\equiv {T_1T_2\over T_1+T_2}$, $T_1$, $T_2$ are the Borel parameters
$a=-(2\pi )^2 \langle {\bar q}q \rangle$.  
We have used the Borel transformation formula:
${\hat {\cal B}}^T_{\omega} e^{\alpha \omega}=\delta (\alpha -{1\over T})$.

Due to the heavy quark symmetry, $\Lambda_{\Sigma}= \Lambda_{\Sigma^*}$
and $f_{\Sigma}= f_{\Sigma^*}$ in the limit $m_Q \to \infty$. So from 
(\ref{final-b-1}) and (\ref{final-b-2}) we have $\eta_2 =\sqrt{3}\eta_1$.
For the decays $\Sigma_c^{*0}\to \Sigma_c^0\gamma$ and 
$\Sigma_c^{*++}\to \Sigma_c^{++}\gamma$, we need make replacement
$(e_u+e_d) \to 2e_d, 2e_u$ in (\ref{final-b-3}).

%%%%%%%%%%%%%%%%%%%%%%%%%%%%%%%%%%%%%%%%%%%%%%%%%%%%%%%%%%%%%%%%%%%%%%%%%%%%%%%%
\section{Radiative decays of heavy mesons}
\label{sec3}
We shall confine ourselves to the lowest lying three doublets and consider
all possible radiative decay processes among them in the leading 
order of $1/m_Q$ expansion. Denote the doublet 
$(1^+,2^+)$ with $j_{\ell}=3/2$ by $(B_1,B_2^*)$, the doublet $(0^+,1^+)$
with $j_{\ell}=1/2$ by $(B^{\prime}_0,B^{\prime}_1)$ and the doublet 
$(0^-, 1^-)$ by $(B, B^*)$. 

The interpolating currents are given in \cite{huang} as 
\begin{eqnarray}
\label{curr1}
&&J^{\dag\alpha}_{1,+,{3\over 2}}=\sqrt{\frac{3}{4}}\:\bar h_v\gamma^5(-i)\left(
{\cal D}_t^{\alpha}-\frac{1}{3}\gamma_t^{\alpha} {\cal D}_t\right)q\;,\\
\label{curr2}
&&J^{\dag\alpha_1,\alpha_2}_{2,+,{3\over 2}}=\sqrt{\frac{1}{2}}\:\bar h_v
\frac{(-i)}{2}\left(\gamma_t^{\alpha_1}{\cal D}_t^{\alpha_2}+
\gamma_t^{\alpha_2}{\cal D}_t^{\alpha_1}-{2\over 3}g_t^{\alpha_1\alpha_2}
 {\cal D}_t\right)q\;,\\
\label{curr3}
&&J^{\dag\alpha}_{1,-,{1\over 2}}=\sqrt{\frac{1}{2}}\:\bar h_v\gamma_t^{\alpha}
q\;,\hspace{1.5cm} J^{\dag\alpha}_{0,-,{1\over 2}}=\sqrt{\frac{1}{2}}\:\bar h_v\gamma_5q\;,
\end{eqnarray}
\begin{eqnarray}
\label{current1}
J^{\dag}_{0,+,{1\over 2}}=\frac{1}{\sqrt{2}}\:\bar h_vq\;,\hspace{1.5cm}
J^{\dag\alpha}_{1,+,{1\over 2}}=\frac{1}{\sqrt{2}}\:\bar h_v\gamma^5\gamma^{\alpha}_tq\;.
\end{eqnarray} 

\begin{itemize}
\item $(1^+, 2^+)\to (0^-, 1^-) + \gamma $

The decay amplitudes are
\begin{eqnarray}\nonumber 
\label{1-1}
 M(B_1\to B^*\gamma  )= e_q e e^{\ast\mu}v^\sigma
\epsilon^*_{\beta}\eta_{\alpha}\{
[\epsilon_{\mu\nu\beta\sigma} q_t^\alpha +(\alpha \leftrightarrow \beta)]
 q_t^\nu  g^1_D(B_1,B^*) & \\
+[\epsilon_{\mu\nu\beta\sigma} (q_t^\alpha q_t^\nu -{1\over 3}q_t^2 g_t^{\alpha\nu})
-(\alpha \leftrightarrow \beta)]g^2_D(B_1,B^*)
+\epsilon_{\mu\alpha\beta\sigma} g_S(B_1,B^*) \}
&\;,
\end{eqnarray}
where the tensor structure associated with 
$g^1_D(B_1,B^*)$ and $g^2_D(B_1,B^*)$ is symmetric and antisymmetric 
under the exchange of $(\alpha \leftrightarrow \beta)$ respectively.
\begin{equation} 
\label{1-2}
 M(B_1\to B\gamma  )= e_q e
e^*_{\beta}\eta_{\alpha}\{
(q_t^\alpha q_t^\beta -{1\over 3}q_t^2 g_t^{\alpha\beta}) g_D(B_1,B)
+g_t^{\alpha\beta} g_S(B_1,B) \}\;,
\end{equation}
\begin{equation} 
\label{1-3}
 M(B_2^*\to B\gamma  )= e_q e
e^*_{\beta}\eta_{\alpha_1\alpha_2}q_\nu v_\sigma 
[ \epsilon^{\beta\nu{\alpha_1}\sigma}q_t^{\alpha_2}
+(\alpha_1 \leftrightarrow \alpha_2 )] g_D (B_2^*,B) \;,
\end{equation}
\begin{eqnarray}
\label{1-4}\nonumber
M(B_2^*\to B^*\gamma  )=e_q e \eta_{\alpha_1 \alpha_2} \eta^*_\beta \{
e_t^\beta (q_t^{\alpha_1}q_t^{\alpha_2}
-{1\over 3} q_t^2 g_t^{\alpha_1 \alpha_2} )g_D^1 (B^*_2, B^*) 
+[e_t^{\alpha_1} (q_t^{\beta}q_t^{\alpha_2} &\\ \nonumber
-{1\over 3} q_t^2 g_t^{\beta \alpha_2} ) 
+(\alpha_1 \leftrightarrow \alpha_2 )] g^2_D (B_2^*,B^*)
+[e_t^{\alpha_1} g_t^{\alpha_2 \beta} 
+(\alpha_1 \leftrightarrow \alpha_2 )] g_S (B_2^*,B^*) \}
& \;,
\end{eqnarray} 
where $\eta_{\mu\nu}$, $\eta_{\mu}$ and $\epsilon_{\mu}$ are polarization
tensors for states $2^+$, $1^+$ and $1^-$ respectively and $e_q e$ is the 
light quark electric charge. 

Due to heavy quark symmetry, there exist only two independent coupling
constants for the D-wave and S-wave decay respectively. Let 
$g_d \equiv g_D (B_2^*,B)$ and $g_s\equiv -g_S (B_2^*,B^*)$. Then we have:
\begin{equation}\label{relat-1}
{\sqrt{6}\over 2}g_S(B_1, B^*)={\sqrt{6}\over 4}g_S(B_1, B)=g_s \; ,
\end{equation}
\begin{eqnarray}\label{relat-2}\nonumber
{\sqrt{6}\over 3}g^1_D(B_1, B^*)=\sqrt{6}g^2_D(B_1, B^*)=
{\sqrt{6}\over 2}g_D(B_1, B) &\\
=-{1\over 2}g_D^1(B_2^*, B^*)
=g_D^2(B_2^*, B^*)=g_d &\; .
\end{eqnarray}
The above relation is confirmed by our detailed calculation. 

For deriving the sum rules for the coupling constants we consider the correlator 
\begin{eqnarray}\nonumber
\label{7a}
 \int d^4x\;e^{-ik\cdot x}\langle\gamma (q)|T\left(J_{0,-,\frac{1}{2}}(0)
 J^{\dagger\alpha}_{1,+,\frac{3}{2}}(x)\right)|0\rangle &\\
 = e_q e
\{ e^*_{\beta} (q_t^\alpha q_t^\beta -{1\over 3}q_t^2 g_t^{\alpha\beta}) 
G^D_{B_1B} (\omega,\omega')
+e_t^{\ast\alpha} G^S_{B_1B} (\omega,\omega') \}&\;. 
\end{eqnarray}

The functions $G^{D,S}_{B_1B}(\omega,\omega^{\prime})$ in (\ref{7a})
have the following double dispersion relation 
\begin{eqnarray}
\label{pole-m-1}
{f_{-,{1\over 2}}f_{+,{3\over 2}}g_{D,S}(B_1B)\over (2\bar\Lambda_{-,{1\over 2}}
-\omega')(2\bar\Lambda_{+,{3\over 2}}-\omega)}+{c\over 2\bar\Lambda_{-,{1\over 2}}
-\omega'}+{c'\over 2\bar\Lambda_{+,{3\over 2}}-\omega}\;,
\end{eqnarray}
where $\bar\Lambda_{P,j_\ell}=m_{P,j_\ell}-m_Q$ and 
$f_{P,j_\ell}$ are constants defined as:
\begin{equation}
\langle 0|J_{j,P,j_{\ell}}^{\alpha_1\cdots\alpha_j}(0)|j',P',j_{\ell}^{'}\rangle=
f_{Pj_l}\delta_{jj'}
\delta_{PP'}\delta_{j_{\ell}j_{\ell}^{'}}\eta^{\alpha_1\cdots\alpha_j}\;.
\end{equation}

Applying the same procedure as in section \ref{sec2} we obtain
\begin{eqnarray}\label{0-d-wave}\nonumber
G^D_{B_1 B}(\omega, \omega')= 
-{\sqrt{6}\over 24} \int_0^{\infty} dt \int_0^1 du e^{i (1-u) {\omega t \over 2}}
e^{i u {\omega' t \over 2}} u \{ 
-{it\over 4}f \psi (u) &\\
+\langle {\bar q} q\rangle [\chi \phi (u) 
+t^2 \left( g_1(u) -g_2(u)\right) ] 
\}+\cdots &\; ,
\end{eqnarray}
\begin{equation}\label{0-s-wave}
G^S_{B_1 B}(\omega, \omega')= -{2\over 3}q_t^2 G^D_{B_1 B}(\omega, \omega')
={1\over 6}(\omega -\omega^\prime )^2 G^D_{B_1 B}(\omega, \omega')\; .
\end{equation}

Finally we have:
\begin{equation}\label{m-0-d}
 g_d f_{-,{1\over 2} } f_{+, {3\over 2} } = {1\over 8}
 e^{ { \Lambda_{-,{1\over 2} } +\Lambda_{+,{3\over 2} } \over T }}
 \{ f u_0 \psi (u_0) +{a\over 2\pi^2} u_0 [\chi \phi (u_0) Tf_0({\omega_c\over T})
 -{4\over T} \left( g_1(u_0)-g_2(u_0)\right)]
 \}\;,
\end{equation}
\begin{eqnarray}\label{m-0-s}\nonumber
 g_s f_{-,{1\over 2} } f_{+, {3\over 2} } = -{1\over 96}
 e^{ { \Lambda_{-,{1\over 2} } +\Lambda_{+,{3\over 2} } \over T }}
 \{ f {d^2\left(u \psi (u)\right)\over du^2}T^2 f_1({\omega_c\over T})
  +{a\over 2\pi^2} [\chi {d^2\left(u \phi (u)\right)\over du^2}
  T^3f_2({\omega_c\over T}) &\\
 -4 {d^2\left( \left( ug_1(u)-ug_2(u)\right)\right)\over du^2}
 Tf_0({\omega_c\over T})]  \}|_{u=u_0}&\;,
\end{eqnarray}
Here we have used integration by parts to absorb the factor $(q\cdot v)^2$, 
which leads to the second derivatives in (\ref{m-0-s}). In this way we 
arrive at the simple form after double Borel transformation.

\item $(0^+, 1^+)\to (0^-, 1^-) + \gamma $

There exists only one independent coupling constant, corresponding to S-wave
decay. The decay amplitudes are:
\begin{equation}
M(B'_1 \to B^* \gamma) =e_q e \epsilon^{\mu\sigma\alpha\beta} e_\mu v_\sigma 
\eta'_\alpha \epsilon^*_\beta g_S (B'_1, B^*) \; ,
\end{equation}
where $\eta'_\alpha$ is the polarization vector of $B'_1$.
\begin{equation}
M(B'_1 \to B \gamma) =e_q e e^\alpha \eta'_\alpha  g_S (B'_1, B) \; ,
\end{equation}
\begin{equation}
M(B'_0 \to B^* \gamma) =e_q e e_t^\beta \epsilon_\beta  g_S (B'_0, B^*) \; .
\end{equation}
The process $B'_0\to B \gamma$ is forbidden due to parity and angular momentum
conservation. Due to heavy quark symmetry, we have
\begin{equation}
g_S (B'_1, B^*)=g_S (B'_1, B)=-g_S (B'_0, B^*)\equiv g_1 \; .
\end{equation}

We consider the correlator
\begin{equation}
 \int d^4x\;e^{-ik\cdot x}\langle \gamma (q)|T\left(J_{0,+,\frac{1}{2}}(0)
 J^{\dagger\beta}_{1,-,\frac{1}{2}}(x)\right)|0\rangle=e_q e e^\beta
 G_{B_0^{\prime} B^*} (\omega,\omega')\;,
\end{equation}
where
\begin{equation}\label{d-wave2}
G_{B_0^{\prime} B^*}(\omega, \omega')= 
-{1\over 4} \langle {\bar q} q\rangle (q\cdot v)
\int_0^{\infty} dt \int_0^1 du e^{i (1-u) {\omega t \over 2}}
e^{i u {\omega' t \over 2}} 
 \{ \chi \phi (u) +t^2  g_1(u) \}+\cdots\; .
\end{equation}

\begin{equation}\label{m-1}
 g_1 f_{-,{1\over 2} } f_{+, {1\over 2} } = {a\over 16\pi^2}
 e^{ { \Lambda_{-,{1\over 2} } +\Lambda_{+,{1\over 2} } \over T }}
 \{ \chi {d\phi (u)\over du} T^2 f_1({\omega_c\over T})
 -4 {dg_1(u) \over du} \}|_{u=u_0}\; .
\end{equation}

\item $(1^+, 2^+)\to (0^+, 1^+) + \gamma $

There exists only one independent coupling constant, corresponding to P-wave
decay. The decay amplitudes are:
\begin{eqnarray}\nonumber
M(B_1\to B'_1 \gamma)=e_q e \eta_\alpha {\eta'}^*_\beta e^*_\mu 
\{ [(q_t^\alpha g_t^{\beta\mu} -{1\over 3} q_t^\mu g_t^{\alpha\beta})
+(\alpha \leftrightarrow \beta )]g_P^1 (B_1, B'_1) &\\
+(q_t^\alpha g_t^{\beta\mu} -q_t^\beta g_t^{\alpha\mu})g_P^2 (B_1, B'_1)
\} &\; ,
\end{eqnarray}
\begin{equation}
M(B_1\to B'_0 \gamma)=e_q e \epsilon_{\mu\sigma\nu\alpha} 
\eta^\alpha  {e^*}^\mu v^\sigma q^\nu g_P (B_1, B'_0) \; ,
\end{equation}
\begin{equation}
M(B_2^*\to B'_0 \gamma)=e_q e {e^*}^\mu \eta_{\alpha_1 \alpha_2}
[(q_t^{\alpha_1} g_t^{\alpha_2 \mu} -{1\over 3} q_t^\mu g_t^{\alpha_1 \alpha_2})
+(\alpha_1 \leftrightarrow \alpha_2 )]g_P (B_2^*, B'_0) \; ,
\end{equation}
\begin{equation}
M(B_2^*\to B'_1 \gamma)=e_q e 
\epsilon_{\mu\sigma\rho\beta} \eta'^\beta 
{e^*}^\mu \eta_{\alpha_1 \alpha_2}
[(q_t^{\alpha_1} g_t^{\alpha_2 \rho} -{1\over 3} q_t^\rho g_t^{\alpha_1 \alpha_2})
+(\alpha_1 \leftrightarrow \alpha_2 )]g_P (B_2^*, B'_1) \; .
\end{equation}

Due to heavy quark symmetry we have
\begin{equation}
{\sqrt{6}\over 3}g_P^1 (B_1, B'_1)=\sqrt{6} g_P^2 (B_1, B'_1)
=\sqrt{6} g_P (B_1, B'_0)=g_P (B_2^*, B'_0)=g_P (B_2^*, B'_1) 
\equiv g_2 \; .
\end{equation}

We consider the correlator
\begin{eqnarray}\nonumber
\int d^4x\;e^{-ik\cdot x}\langle\pi(q)|T\left(J^{\alpha}_{0,+,\frac{1}{2}}(0)
 J^{\dagger\alpha_1\alpha_2}_{2,+,\frac{3}{2}}(x)\right)|0\rangle= & \\
 e_q e e^\mu 
[(q_t^{\alpha_1} g_t^{\alpha_2 \mu} -{1\over 3} q_t^\mu g_t^{\alpha_1 \alpha_2})
+(\alpha_1 \leftrightarrow \alpha_2 )]
G_{B_2^* B'_0} (\omega,\omega') &\;,
\end{eqnarray}
where
\begin{equation}
G_{B_2^* B'_0}(\omega, \omega')= 
-{1\over 8} \langle {\bar q} q\rangle (q\cdot v)
\int_0^{\infty} dt \int_0^1 du e^{i (1-u) {\omega t \over 2}}
e^{i u {\omega' t \over 2}} 
 u \{ \chi \phi (u) +t^2  g_1(u) \}+\cdots\; .
\end{equation}

\begin{equation}\label{m-2}
 g_2 f_{+,{1\over 2} } f_{+, {3\over 2} } = {a\over 32\pi^2}
 e^{ { \Lambda_{+,{1\over 2} } +\Lambda_{+,{3\over 2} } \over T }}
 \{ \chi {d\left( u\phi (u)\right)\over du} T^2 f_1({\omega_c\over T})
 -4 {d\left( u g_1(u) \right)\over du} \}|_{u=u_0}\; .
\end{equation}

\item $B'_1 \to B'_0 \gamma$

\begin{equation}
M(B'_1 \to B'_0 \gamma )=e_q e \epsilon^{\alpha\mu\nu\sigma}
\eta'_\alpha e^*_\mu q_\nu v_\sigma g_3 \; .
\end{equation}

In order to derive $g_3$, we consider the correlator
\begin{equation}
\int d^4x\;e^{-ik\cdot x}\langle\gamma   (q)|T\left(J_{0,+,\frac{1}{2}}(0)
 J^{\dagger\alpha}_{1,+,\frac{1}{2}}(x)\right)|0\rangle=e_q e
 \epsilon^{\alpha\mu\nu\sigma} e_\mu q_\nu v_\sigma
 G_{B_1^{\prime}B_0^{\prime}} (\omega,\omega')\;,
\end{equation}
where 
\begin{eqnarray}\nonumber
G_{B'_1 B'_0}(\omega, \omega')= 
{i\over 4} \int_0^{\infty} dt \int_0^1 du e^{i (1-u) {\omega t \over 2}}
e^{i u {\omega' t \over 2}}  \{ 
{it\over 4}f \psi (u) +\langle {\bar q} q\rangle [\chi \phi (u) &\\
+t^2 \left( g_1(u) -g_2(u)\right) ] 
\}+\cdots &\; .
\end{eqnarray}

\begin{equation}\label{m-3}
 g_3 f^2_{+,{1\over 2} }= -{1\over 4}
 e^{ { 2\Lambda_{+,{1\over 2} } \over T }}
 \{ - f  \psi (u_0) +{a\over 2\pi^2} [\chi \phi (u_0) Tf_0({\omega_c\over T})
 -{4\over T} \left( g_1(u_0)-g_2(u_0)\right)]
 \}\; .
\end{equation}

\item $B^* \to B \gamma$

\begin{equation}
M(B^* \to B \gamma )=e_q e \epsilon^{\alpha\mu\nu\sigma}
\epsilon_\alpha e^*_\mu q_\nu v_\sigma g_4 \; .
\end{equation}

In order to derive $g_4$, we consider the correlator
\begin{equation}
\int d^4x\;e^{-ik\cdot x}\langle\gamma   (q)|T\left(J_{0,-,\frac{1}{2}}(0)
 J^{\dagger\alpha}_{1,-,\frac{1}{2}}(x)\right)|0\rangle=e_q e
 \epsilon^{\alpha\mu\nu\sigma} e_\mu q_\nu v_\sigma
 G_{B^* B} (\omega,\omega')\;,
\end{equation}
where 
\begin{eqnarray}\nonumber
G_{B^* B}(\omega, \omega')= 
{i\over 4} \int_0^{\infty} dt \int_0^1 du e^{i (1-u) {\omega t \over 2}}
e^{i u {\omega' t \over 2}}  \{ 
-{it\over 4}f \psi (u) +\langle {\bar q} q\rangle [\chi \phi (u) &\\
+t^2 \left( g_1(u) -g_2(u)\right) ] 
\}+\cdots &\; .
\end{eqnarray}
Note this coupling was calculated in \cite{aliev} using LCQSR. But there the 
contribution from the photon wave function $\psi (u)$ has not been taken 
into account.

\begin{equation}\label{m-4}
 g_4 f^2_{-,{1\over 2} }= -{1\over 4}
 e^{ { 2\Lambda_{-,{1\over 2} } \over T }}
 \{  f  \psi (u_0) +{a\over 2\pi^2} [\chi \phi (u_0) Tf_0({\omega_c\over T})
 -{4\over T} \left( g_1(u_0)-g_2(u_0)\right)]
 \}\; .
\end{equation}

\item $B_2^* \to B_1 \gamma$
\begin{equation}
M(B_2^* \to B_1 \gamma )=e_q e \epsilon^{\alpha\mu\nu\sigma}
 e_\mu q_\nu v_\sigma \eta^*_\beta \eta_{\alpha_1 \alpha_2}
 (g_t^{\alpha_1 \rho} q_t^{\alpha_2}+g_t^{\alpha_2 \rho} q_t^{\alpha_1}
-{2\over 3} g_t^{\alpha_1 \alpha_2} q_t^{\rho} )
(2 q_t^\beta g_t^{\rho \alpha} +g_t^{\alpha\beta}q_t^\rho )
g_5 \; .
\end{equation}

In order to derive $g_5$, we consider the correlator
\begin{eqnarray}\nonumber
\int d^4x\;e^{-ik\cdot x}\langle\gamma  (q)|T\left(J^{\beta}_{1,+,\frac{3}{2}}(0)
 J^{\dagger\alpha_1 \alpha_2}_{2,+,\frac{3}{2}}(x)\right)|0\rangle= &\\  
e_q e \epsilon^{\alpha\mu\nu\sigma}  e_\mu q_\nu v_\sigma 
 (g_t^{\alpha_1 \rho} q_t^{\alpha_2}+g_t^{\alpha_2 \rho} q_t^{\alpha_1}
-{2\over 3} g_t^{\alpha_1 \alpha_2} q_t^{\rho} )
(2 q_t^\beta g_t^{\rho \alpha} +g_t^{\alpha\beta}q_t^\rho )
 G_{B_2^*B_1} (\omega,\omega') & \;,
\end{eqnarray}
where 
\begin{eqnarray}\nonumber
G_{B_2^* B_1}(\omega, \omega')= 
-i{\sqrt{6}\over 16} \int_0^{\infty} dt \int_0^1 du e^{i (1-u) {\omega t \over 2}}
e^{i u {\omega' t \over 2}} u(1-u) \{ 
-{it\over 4}f \psi (u)&\\
+\langle {\bar q} q\rangle [\chi \phi (u) 
+t^2 \left( g_1(u) -g_2(u)\right) ] 
\}+\cdots &\; .
\end{eqnarray}

\begin{equation}\label{m-5}
 g_5 f^2_{+,{3\over 2} }= {\sqrt{6}\over 16}
 e^{ { 2\Lambda_{+,{3\over 2} } \over T }}
u_0 (1-u_0) \{  f  \psi (u_0) +{a\over 2\pi^2} [\chi \phi (u_0) Tf_0({\omega_c\over T})
 -{4\over T} \left( g_1(u_0)-g_2(u_0)\right)]
 \}\; .
\end{equation}

\end{itemize}

%%%%%%%%%%%%%%%%%%%%%%%%%%%%%%%%%%%%%%%%%%%%%%%%%%%%%%%%%%%%%%%%%%%%%%%%%%%%%%%%
\section{Determination of the parameters}
\label{sec4} 

The leading photon wave functions receive only small corrections from 
the higher conformal spins \cite{braun-lc} so they do not deviate 
much from the asymptotic form. We shall use \cite{ali}
\begin{equation}
\phi (u) =6u{\bar u} \; ,
\end{equation}
\begin{equation}
\psi (u) =1 \; ,
\end{equation}
\begin{equation}
g_1 (u) =-{1\over 8}{\bar u} (3-u)\; ,
\end{equation}
\begin{equation}
g_2 (u) =-{1\over 4}{\bar u}^2 \; .
\end{equation}
with $f=0.028$GeV$^2$ and $\chi =-4.4 $GeV$^2$ \cite{kogan} at the scale $\mu =1$GeV. 
Using this value of $\chi$, the octet, decuplet and heavy baryon magnetic 
moments have been calculated to a good accuracy \cite{ioffe-mag,chiu,zhu-mag}. 

We need the mass parameters $\bar\Lambda$'s and the coupling constants $f$'s of the
corresponding interpolating currents in the leading order of $\alpha_s$ 
as input. The results are \cite{mass,zhu}
\begin{eqnarray}\nonumber
\label{fvalue-b}
\bar\Lambda_{\Lambda}=0.8 ~~\mbox{GeV}\hspace{1.2cm}f_{\Lambda}
=(0.018\pm 0.002) ~~\mbox{GeV}^3 \;, &\\
\bar\Lambda_{\Sigma}=1.0 ~~\mbox{GeV}\hspace{1.1cm}f_{\Sigma}
=(0.04\pm 0.004)  ~~\mbox{GeV}^3 &\;.
\end{eqnarray}
\begin{eqnarray}\nonumber
\label{fvalue-m}
\bar\Lambda_{+,3/2}=0.82 ~~\mbox{GeV}
\hspace{1.2cm}f_{+,3/2}=0.19\pm 0.03 ~~\mbox{GeV}^{5/2}\;, &\\ \nonumber
\bar\Lambda_{+,1/2}=1.1 ~~\mbox{GeV}
\hspace{1.1cm}f_{+,1/2}=0.40\pm 0.06 ~~\mbox{GeV}^{3/2}\;, &\\
\bar\Lambda_{-,1/2}=0.5 ~~\mbox{GeV}
\hspace{1.1cm}f_{-,1/2}=0.25 ~~\mbox{GeV}^{3/2}& \;.
\end{eqnarray}

We choose to work at the symmetric point $T_1=T_2=2T$, i.e., 
$u_0 ={1\over 2}$. Such a choice is very reasonable for the symmetric sum rules 
(\ref{final-b-3}), (\ref{m-3}), (\ref{m-4})
and (\ref{m-5}) since $\Sigma_c^*$ and $\Sigma_c$, and the three meson 
doublets are degenerate in the leading order of HQET. 
Moreover, the mass difference between $\Sigma_c^*$ and $\Lambda_c$ is 
only about $0.2$GeV. The $(0^+, 1^+)$ doublet lies only slightly below
$(1^+, 2^+)$ doublet. Due to the large values of $T_1$, $T_2$
used below, the choice of $T_1=T_2$ is 
also reasonable for sum rules (\ref{final-b-1}) and (\ref{m-2}).

Note the choice $T_1 =T_2$ is not unique for the asymmetric sum rules
(\ref{m-0-d}), (\ref{m-0-s}) and (\ref{m-1}) since the initial and final 
mesons have different masses. But the choice $T_1 =T_2$ will enable 
the clean subtraction of the continuum contribution, which is cruicial 
for the numerical analysis of the sum rules.  
In our case the sum rules are stable with reasonable variations
of the Borel parameter $T_1$ and $T_2$.  
Such a choice does not alter significantly the numerical results.
Based on these considerations we adopt $u_0={1\over 2}$ 
for for the sum rules (\ref{m-0-d}), (\ref{m-0-s}) and (\ref{m-1})
too.

%%%%%%%%%%%%%%%%%%%%%%%%%%%%%%%%%%%%%%%%%%%%%%%%%%%%%%%%%%%%%%%%%%%%%%%%%%%%%%%
\section{Numerical results and discussion}
\label{sec5}

\subsection{Numerical analysis of the baryon sum rules}

We now turn to the numerical evaluation of the sum rules for the coupling
constants. 
Since the spectral density of the sum rule (\ref{final-b-1})-(\ref{final-b-3}) 
$\rho (s)$ is either proptional to $s^2$ or $s^3$, the continuum has to be 
subtracted carefully. We use the value of the continuum threshold $\omega_c$ 
determined from the corresponding mass sum rule at the leading order of 
$\alpha_s$ and $1/m_Q$ \cite{mass}.

The lower limit of $T$ is 
determined by the requirement that the terms of higher
twists in the operator expansion is reasonably smaller than the
leading twist, say $\leq 1/3$ of the latter. This leads to $T>1.3$ GeV for
the sum rules (\ref{final-b-1})-(\ref{final-b-3}).
In fact the twist-four terms contribute only a few percent to the sum rules. 
The upper limit of $T$ is constrained by the requirement that
the continuum contribution is less than $50\%$. This corresponds to $T<2.2$GeV.

The variation of $\eta_{1,3}$ with the Borel parameter $T$ and $\omega_c$ is 
presented in FIG. 1 and FIG. 2. The curves correspond to 
$\omega_c =2.4, 2.5, 2.6$GeV from bottom to top respectively.
Stability develops for the sum rules (\ref{final-b-1}) and (\ref{final-b-3}) 
in the region $1.3$ GeV $<$$T$$<$$2.2$ GeV, we get: 
\begin{eqnarray}
\label{num-b-1}
 &&\eta_1 f_{\Sigma} f_{\Lambda} =(7.0\pm 0.9)\times 10^{-4}\mbox{GeV}^5\;,\\
 &&\eta_3 f_{\Sigma^*} f_{\Sigma}=(3.9\pm 0.5)\times 10^{-4}\mbox{GeV}^5\;,
\end{eqnarray}
where the errors refers to the variations with $T$ and 
$\omega_c$ in this region. And the central value corresponds 
to $T=1.6$GeV and $\omega_c =2.5$GeV.

Combining (\ref{fvalue-b}) we arrive at 
\begin{eqnarray}
\label{num-b-2}
 &&\eta_1  =(1.0\pm 0.2)\mbox{GeV}^{-1}\;,\\
 &&\eta_3 =(0.24\pm 0.05)\mbox{GeV}^{-1}\;.
\end{eqnarray}

\subsection{Numerical analysis of the meson sum rules}

We now turn to the numerical evaluation of the sum rules for the coupling
constants. The lower limit of $T$ is determined by the requirement 
that the terms of higher twists in the operator expansion is less 
than one third of the whole sum rule. This leads to $T>1.0$ GeV for
the sum rules (\ref{m-0-d}), (\ref{m-0-s}), (\ref{m-1}), (\ref{m-2}), 
(\ref{m-3}), (\ref{m-4}) and (\ref{m-5}).
In fact the twist-four terms contribute only a few percent to the sum rules 
for such $T$ values. The upper limit of $T$ is constrained by the requirement that
the continuum contribution is less than $30\%$. This corresponds to $T<2.5$GeV.
With the values of photon wave functions at $u_0 ={1\over 2}$ we obtain 
the left hand side of these sum rules as functions of $T$. 
The continuum threshold is $\omega_c =3.0 \pm 0.2$GeV except  
$\omega_c =2.4 \pm 0.2$GeV for the sum rule (\ref{m-4}).
Stability develops for the sum rules in the region $1.0$ GeV $<$$T$$<$$2.5$ GeV. 
The results are shown in FIG. 3-9.
Numerically we have:
\begin{eqnarray}
\label{res}
 &&g_df_{-,{1\over 2} } f_{+, {3\over 2} }
   =-(3.0\pm 0.2)\times 10^{-2}~~~\mbox{GeV}^{2}\;,\\
 &&g_sf_{-,{1\over 2} } f_{+, {3\over 2} }
   =-(1.9\pm  0.2)\times 10^{-2}~~~\mbox{GeV}^{4}\;,\\
 &&g_1f_{-,{1\over 2} } f_{+, {1\over 2} }
   =-(1.5\pm  0.5)\times 10^{-2}~~~\mbox{GeV}^{3}\;,\\
 &&g_2f_{+,{1\over 2} } f_{+, {3\over 2} }
   =-(5.5\pm  0.4)\times 10^{-2}~~~\mbox{GeV}^{3}\;,\\
 &&g_3f^2_{+,{1\over 2} }
   =(0.28\pm  0.04)~~~\mbox{GeV}^{2}\;,\\
 &&g_4f^2_{-,{1\over 2} }
   =(8.9\pm  0.5)\times 10^{-2}~~~\mbox{GeV}^{2}\;,\\
 &&g_5f^2_{+,{3\over 2} } 
   =-(2.3\pm  0.3)\times 10^{-2}~~~\mbox{GeV}^{2}\;,
\end{eqnarray}
where the errors refer to the variations with $T$  
in this region and the uncertainty in $\omega_c$. And the central value corresponds 
to $T=1.5$GeV and $\omega_c =3.0 \pm 0.2$GeV except that 
we use $\omega_c =2.4 \pm 0.2$GeV for the sum rule (\ref{m-4}).

With the central values of f's in (\ref{fvalue-m}) we get the absolute
value of the coupling constants:
\begin{eqnarray}
\label{result}
 &&g_d=-(0.63\pm 0.10)~~~\mbox{GeV}^{-2}\;,\\
 &&g_s=-(0.40\pm 0.05)\;,\\
 &&g_1=-(0.20\pm 0.06)\;,\\
 &&g_2=-(0.72\pm 0.07)~~~\mbox{GeV}^{-1}\;,\\
 &&g_3=(1.8\pm 0.3)~~~\mbox{GeV}^{-1}\;,\\
 &&g_4=(1.4\pm 0.2)~~~\mbox{GeV}^{-1}\;,\\
 &&g_5=-(0.64\pm 0.08)~~~\mbox{GeV}^{-3}\;.
\end{eqnarray}

Note we have only considered the uncertainty due to the variations of the 
Borel parameter and the continuum threshold in the above expressions. 
There are other sources of uncertainty. The input parameters $\chi$ and $f$
are associated with the photon distribution amplitude. Especially the 
value of $\chi$ has been estimated with QCD sum rules \cite{kogan} and 
with the octet baryon magnetic moments as inputs using the external field
method \cite{chiu}. Both approaches yield consistent results 
$\chi \approx -4.4$ GeV. With this value the octet, decuplet and
heavy baryon magnetic moments derived using the external field method 
are in good agreement with the experimental data. So we expect its accuracy
is better than $30\%$. The value of $f$ has been estimated with the 
vector meson dominance model also with an accuracy of $30\%$ \cite{mendel}.

The light cone sum rules for the coupling constants $g_i$ and the mass 
sum rules for the heavy hadrons in HQET both receive large perturbative
QCD corrections. But their ratio does not depend on radiative corrections
strongly because of large cancellation \cite{report}. In the present case,
the uncertainty of the coupling constants $g_i$ due to radiative 
corrections is expected to around $10\%$ while the couplings $f_{\Lambda}$ 
etc are affected significantly. 

Another possible source of error is the truncation of the light cone 
expansion at twist four. We take the sum rules for $\eta_1$ for example.
At $T=1.5$ GeV, the twist-four term involved with $g_1, g_2$ is only 
$-2.5\%$ of the leading twist term after making double Borel transformation
to (\ref{quark1}). Even after the subtraction of the continuum and excited 
states contribution the twist-four term is only $-15\%$ of the twist-two one
in (\ref{final-b-1}). So the light cone expansion converges quickly. 
We expect the contribution of higher twist distribution amplitudes to 
be small.

We have calculated the coupling constant in the leading order of HQET. 
The $1/m_Q$ correction is sizable for the charm system. But for the 
bottom system the $1/m_Q$ correction is typically around $5\% \sim 10\%$
for the pionic coupling constants \cite{zhu}. We expect the $1/m_Q$ correction
to the electromagnetic coupling constants is of the same order. 
The inherent uncertainty of the method of QCD sum rules is not included,
which is typically about $10\%$.

\subsection{Decay widths of heavy hadrons}

With these coupling constants we can calculate the decay widths of 
heavy hadrons.

The decay width formulas in the leading order of HQET are 
\begin{eqnarray}
\label{widths}
&&\Gamma(\Sigma_b\to \Lambda_b \gamma )=
4\eta_1^2 \alpha |\vec q|^3 \;,\nonumber\\
&&\Gamma(\Sigma^*_b\to \Lambda_b \gamma )=
\eta_2^2 \alpha |\vec q|^3 
{3m_i^2 +m_f^2\over 3m_i^2}\;,\nonumber\\
&&\Gamma(\Sigma^*_b\to \Sigma_b \gamma )=
\eta_3^2 \alpha |\vec q|^3 
{3m_i^2 +m_f^2\over 3m_i^2}\;,\nonumber\\
&&\Gamma(B_1\to B^*\gamma )={2\over 3} e_q^2 \alpha
 ({14\over 9} g_d^2|\vec q|^5 +g_s^2|\vec q|) \;,\nonumber\\
&&\Gamma(B_1\to B\gamma )={4\alpha\over 3} e_q^2 \alpha
 ({1\over 18} g_d^2|\vec q|^5 +g_s^2|\vec q|) \;,\nonumber\\
&&\Gamma(B_2^*\to B\gamma )= 
{2\over 5}e_q^2 \alpha g_d^2|\vec q|^5\;,\nonumber\\
&&\Gamma(B_2^*\to B^*\gamma )=e_q^2 \alpha 
 ({64\over 45} g_d^2|\vec q|^5 +4g_s^2|\vec q|) \;,\nonumber\\
&&\Gamma(B'_1\to B^*\gamma )=e_q^2 \alpha
 g_1^2|\vec q| \;,\nonumber\\
&&\Gamma(B'_1\to B\gamma )={1\over 2} e_q^2 \alpha
 g_1^2|\vec q| \;,\nonumber\\
&&\Gamma(B'_0\to B^*\gamma )={3\over 2} e_q^2 \alpha
 g_1^2|\vec q| \;,\nonumber\\
&&\Gamma(B^*\to B\gamma )={1\over 3} e_q^2 \alpha
 g_4^2|\vec q|^3 \;.
\end{eqnarray}
where $|\vec q|={m_i^2-m_f^2\over 2m_i}$, 
$m_i$, $m_f$ is the parent and decay heavy hadron mass.

We apply the leading order formulas obtained above to the
excited states of bottomed hadrons using the central values 
of the coupling constants in the previous section. 
\begin{eqnarray}
\label{num-bottom}
&&\Gamma(\Sigma_b\to \Lambda_b \gamma )=131
\times \left( {|\vec q|\over 165\mbox{MeV}}\right)^3 ~\mbox{keV}
\;,\nonumber \\
&&\Gamma(\Sigma^{*0}_b\to \Lambda_b \gamma )=313
\times \left( {|\vec q|\over 224\mbox{MeV}}\right)^3 ~\mbox{keV}
\;,\nonumber \\
&&\Gamma(\Sigma^{*+}_b\to \Sigma^+_b \gamma )=2.2
\times \left( {|\vec q|\over 63.4\mbox{MeV}}\right)^3 ~\mbox{keV}
\;,\nonumber \\
&&\Gamma(\Sigma^{*0}_b\to \Sigma^0_b \gamma )=0.14
\times \left( {|\vec q|\over 63.4\mbox{MeV}}\right)^3 ~\mbox{keV}
\;,\nonumber \\
&&\Gamma(\Sigma^{*-}_b\to \Sigma^-_b \gamma )=0.56
\times \left( {|\vec q|\over 63.4\mbox{MeV}}\right)^3 ~\mbox{keV}
\;,\nonumber \\
&&\Gamma(B^{*0}\to B^0\gamma )=1.4
\times \left( {|\vec q|\over 137\mbox{MeV}}\right) ~\mbox{keV}
\;,\nonumber \\
&&\Gamma(B^{*+}\to B^+\gamma )=5.5
\times \left( {|\vec q|\over 137\mbox{MeV}}\right) ~\mbox{keV}
\;,\nonumber \\
&&\Gamma(B^0_1\to B^0\gamma )=84.4
\times \left( {|\vec q|\over 490\mbox{MeV}}\right) ~\mbox{keV}
\;,\nonumber \\
&&\Gamma(B^+_1\to B^+\gamma )=338
\times \left( {|\vec q|\over 490\mbox{MeV}}\right) ~\mbox{keV}
\;,\nonumber \\
&&\Gamma(B^0_1\to B^{*0}\gamma )=42.2
\times \left( {|\vec q|\over 377\mbox{MeV}}\right) ~\mbox{keV}
\;,\nonumber \\
&&\Gamma(B^+_1\to B^{*+}\gamma )=169
\times \left( {|\vec q|\over 377\mbox{MeV}}\right) ~\mbox{keV}
\;,\nonumber \\
&&\Gamma(B_2^{*0}\to B^0\gamma )=5.8
\times \left( {|\vec q|\over 537\mbox{MeV}}\right) ~\mbox{keV}
\;,\nonumber \\
&&\Gamma(B_2^{*+}\to B^+\gamma )=23
\times \left( {|\vec q|\over 537\mbox{MeV}}\right) ~\mbox{keV}
\;,\nonumber \\
&&\Gamma(B_2^{*0}\to B^{*0}\gamma )=211
\times \left( {|\vec q|\over 408\mbox{MeV}}\right) ~\mbox{keV}
\;,\nonumber \\
&&\Gamma(B_2^{*+}\to B^{*+}\gamma )=844
\times \left( {|\vec q|\over 408\mbox{MeV}}\right) ~\mbox{keV}
\;.
\end{eqnarray}
The uncertainty of the decay width is typically about $30\%$.

We do not present numerical results for the radiative decay widths
for the charmed hadrons since $1/m_Q$ corrections are sizable for
the charm system while such corrections are only a few percent 
of the leading order term for the bottom system \cite{zhu}.

In summary we have calculated the coupling constants of photons 
with the heavy baryons and the lowest three heavy meson doublets
using the light cone QCD sum rules with the photon wave functions
in the leading order of HQET. We hope these calculations will be 
tested in the future experiments.

%%%%%%%%%%%%%%%%%%%%%%%%%%%%%%%%%%%%%%%%%%%%%%%%%%%%%%%%%%%%%%%%%%%%%%%%%%%%%%%%

\vspace{0.8cm} {\it Acknowledgments:\/} S.-L. Zhu was supported by
the Postdoctoral Science Foundation of China and 
the Natural Science Foundation of China. Y.D. was supported by 
the Natural Science Foundation of China.
\bigskip
\vspace{1.cm}

{\bf Figure Captions}
\vspace{2ex}
\begin{center}
\begin{minipage}{120mm}
{\sf
FIG. 1.} \small{Dependence of $f_{\Sigma} f_{\Lambda}\eta_1$ on the Borel parameter
 $T$ for different values of the continuum threshold $\omega_c$. 
From top to bottom the curves correspond to $\omega_c=2.6, 2.5, 2.4$ GeV. }
\end{minipage}
\end{center}
\begin{center}
\begin{minipage}{120mm}
{\sf
FIG. 2.} \small{Dependence of $f_{\Sigma^*} f_{\Sigma}\eta_3$ on  
$T$ with $\omega_c=2.6, 2.5, 2.4$ GeV. }
\end{minipage}
\end{center}
\begin{center}
\begin{minipage}{120mm}
{\sf
FIG. 3.} \small{Dependence of $g_df_{-,{1\over 2} } f_{+, {3\over 2} }$ on  
$T$ with $\omega_c=3.2, 3.0, 2.8$ GeV. }
\end{minipage}
\end{center}
\begin{center}
\begin{minipage}{120mm}
{\sf
FIG. 4.} \small{Dependence of $g_sf_{-,{1\over 2} } f_{+, {3\over 2} }$ on  
$T$ with $\omega_c=3.2, 3.0, 2.8$ GeV. }
\end{minipage}
\end{center}
\begin{center}
\begin{minipage}{120mm}
{\sf
FIG. 5.} \small{Dependence of $g_1f_{-,{1\over 2} } f_{+, {1\over 2} }$ on $T$. }
\end{minipage}
\end{center}
\begin{center}
\begin{minipage}{120mm}
{\sf
FIG. 6.} \small{Dependence of $g_2f_{+,{1\over 2} } f_{+, {3\over 2} }$ on  
$T$ with $\omega_c=3.2, 3.0, 2.8$ GeV. }
\end{minipage}
\end{center}
\begin{center}
\begin{minipage}{120mm}
{\sf
FIG. 7.} \small{Dependence of $g_3f^2_{+,{1\over 2} }$ on  
$T$ with $\omega_c=3.2, 3.0, 2.8$ GeV. }
\end{minipage}
\end{center}
\begin{center}
\begin{minipage}{120mm}
{\sf
FIG. 8.} \small{Dependence of $g_4f^2_{-,{1\over 2} }$ on  
$T$ with $\omega_c=2.6, 2.4, 2.2$ GeV. }
\end{minipage}
\end{center}
\begin{center}
\begin{minipage}{120mm}
{\sf
FIG. 9.} \small{Dependence of $g_5f^2_{+,{3\over 2} }$ on  
$T$ with $\omega_c=3.2, 3.0, 2.8$ GeV. }
\end{minipage}
\end{center}

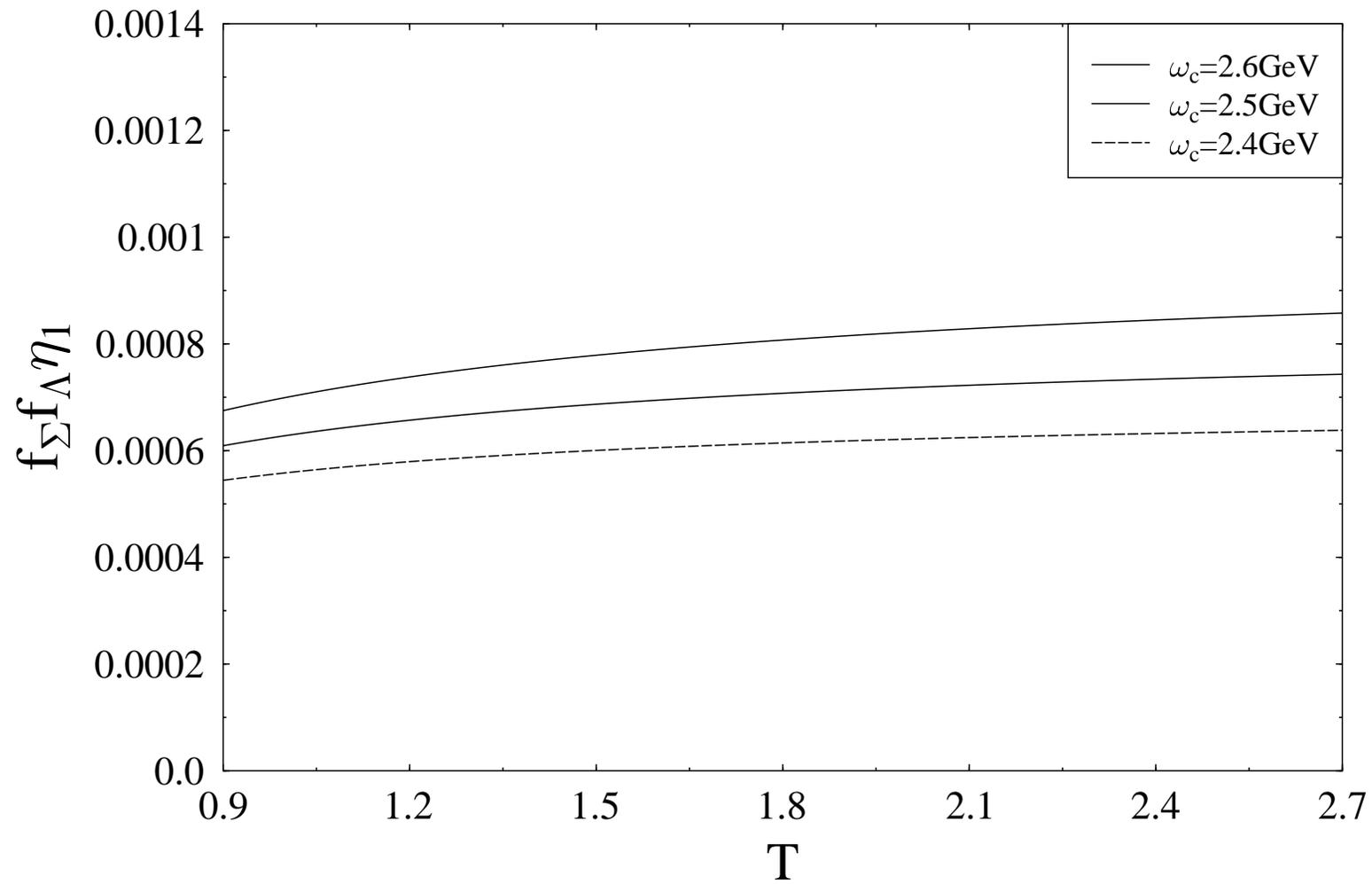
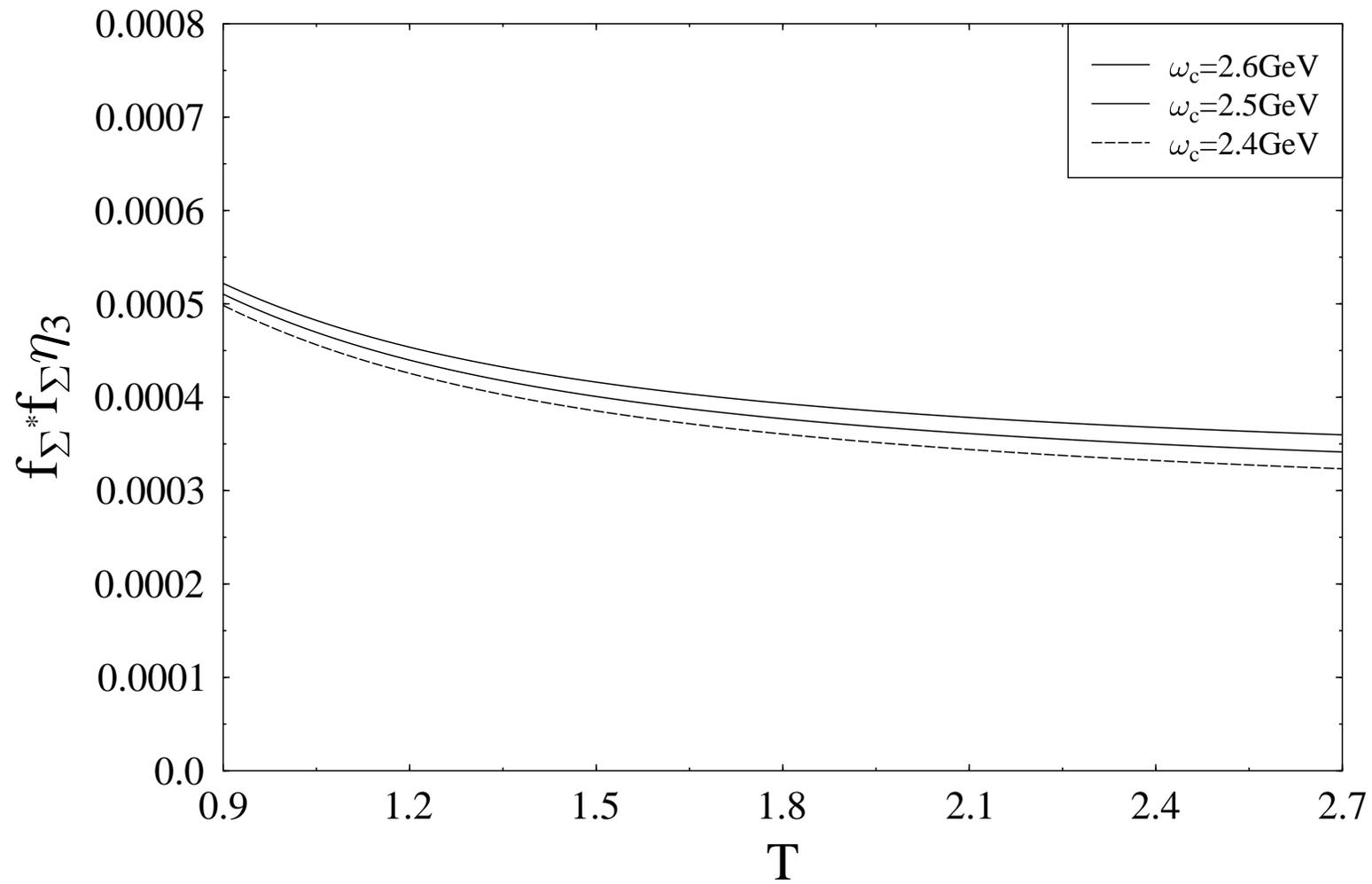
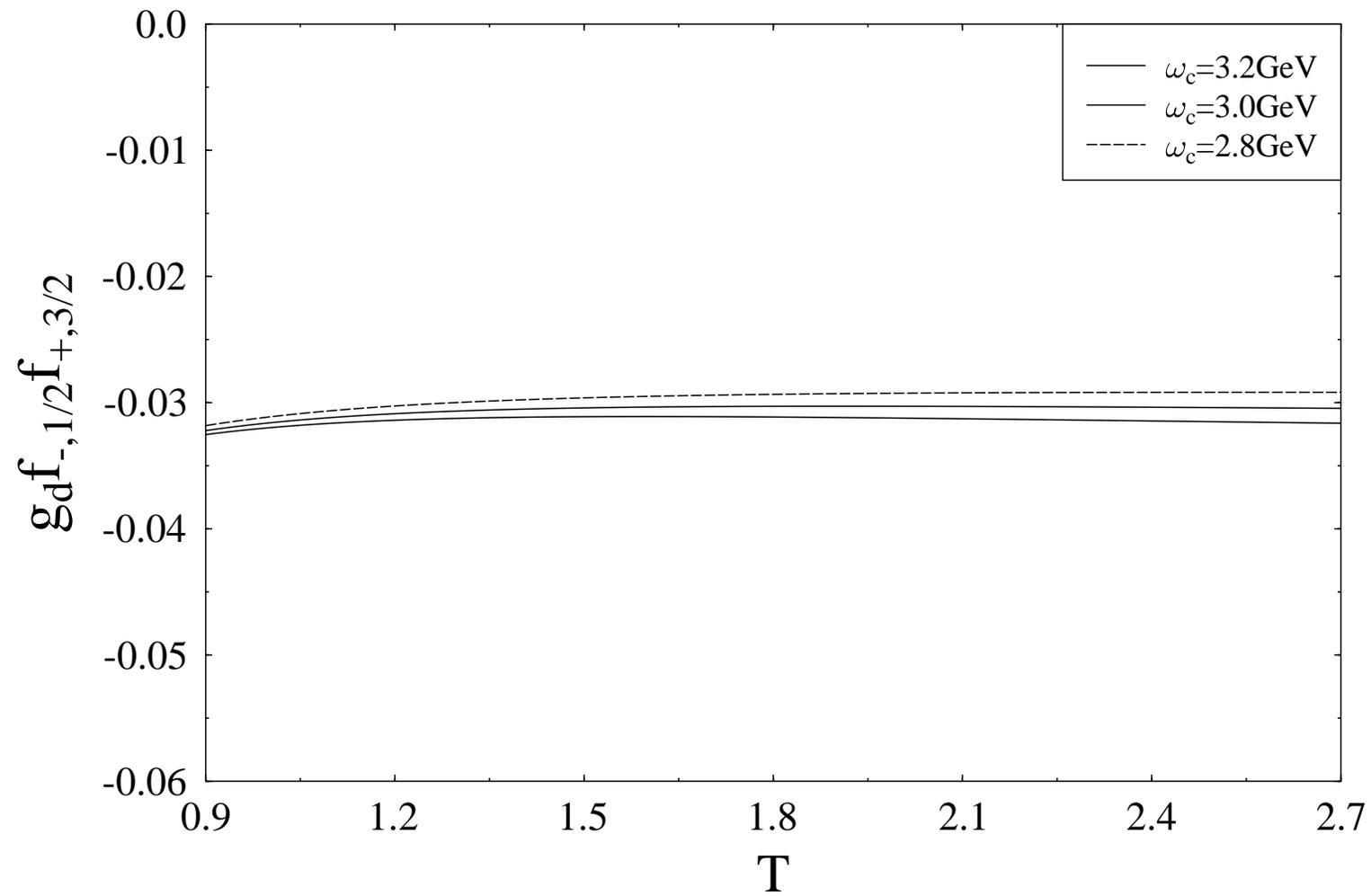
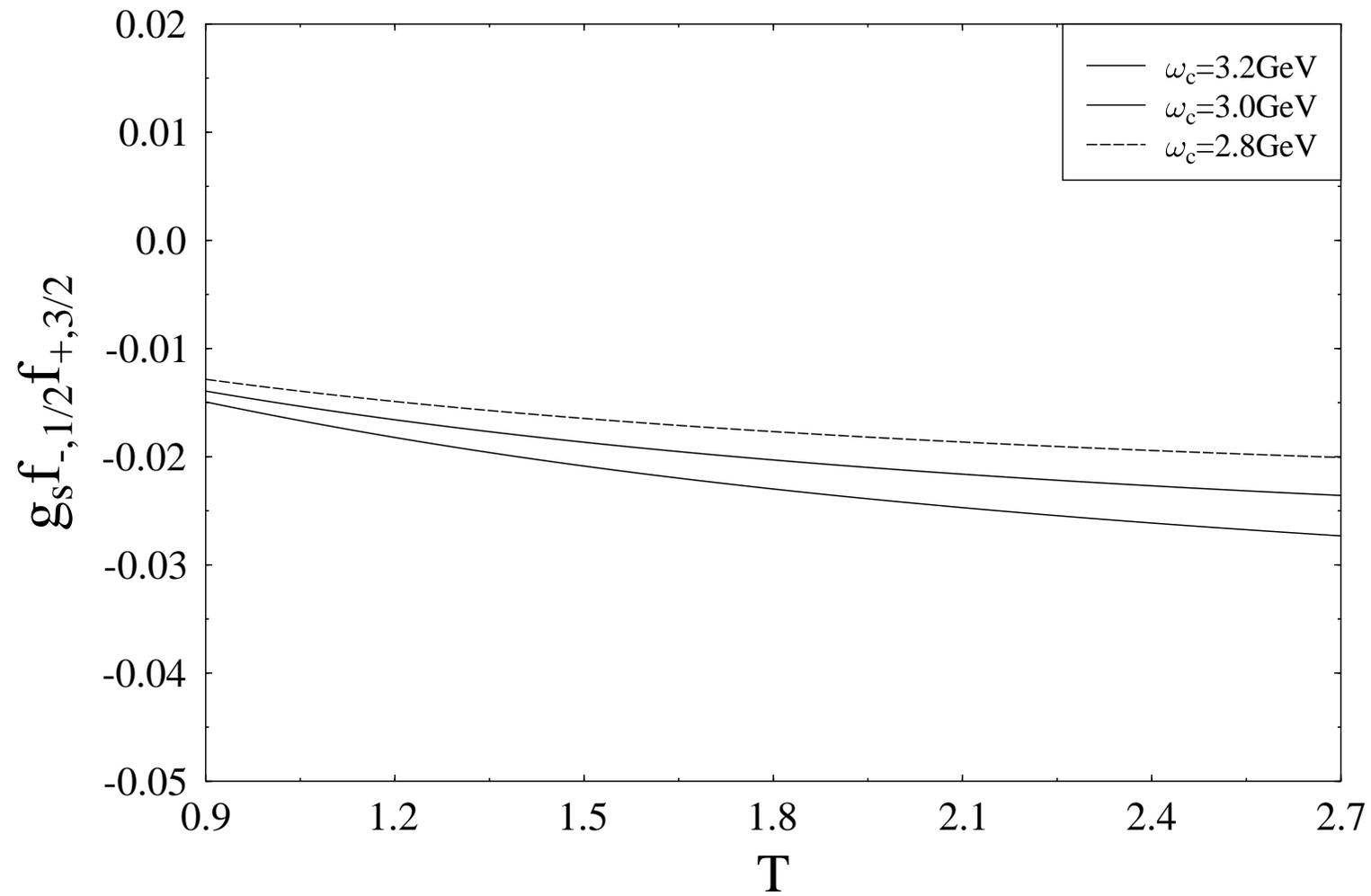
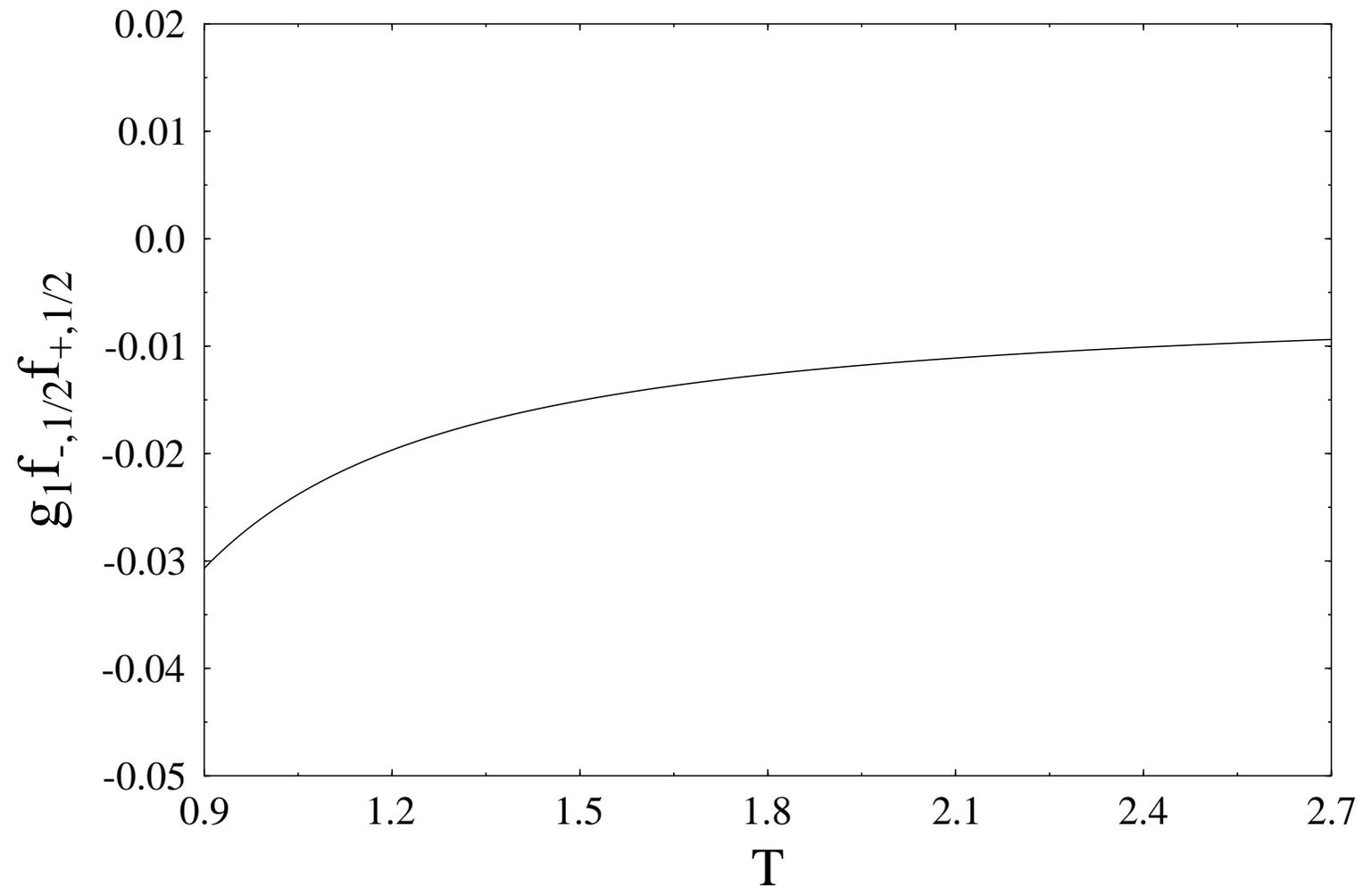
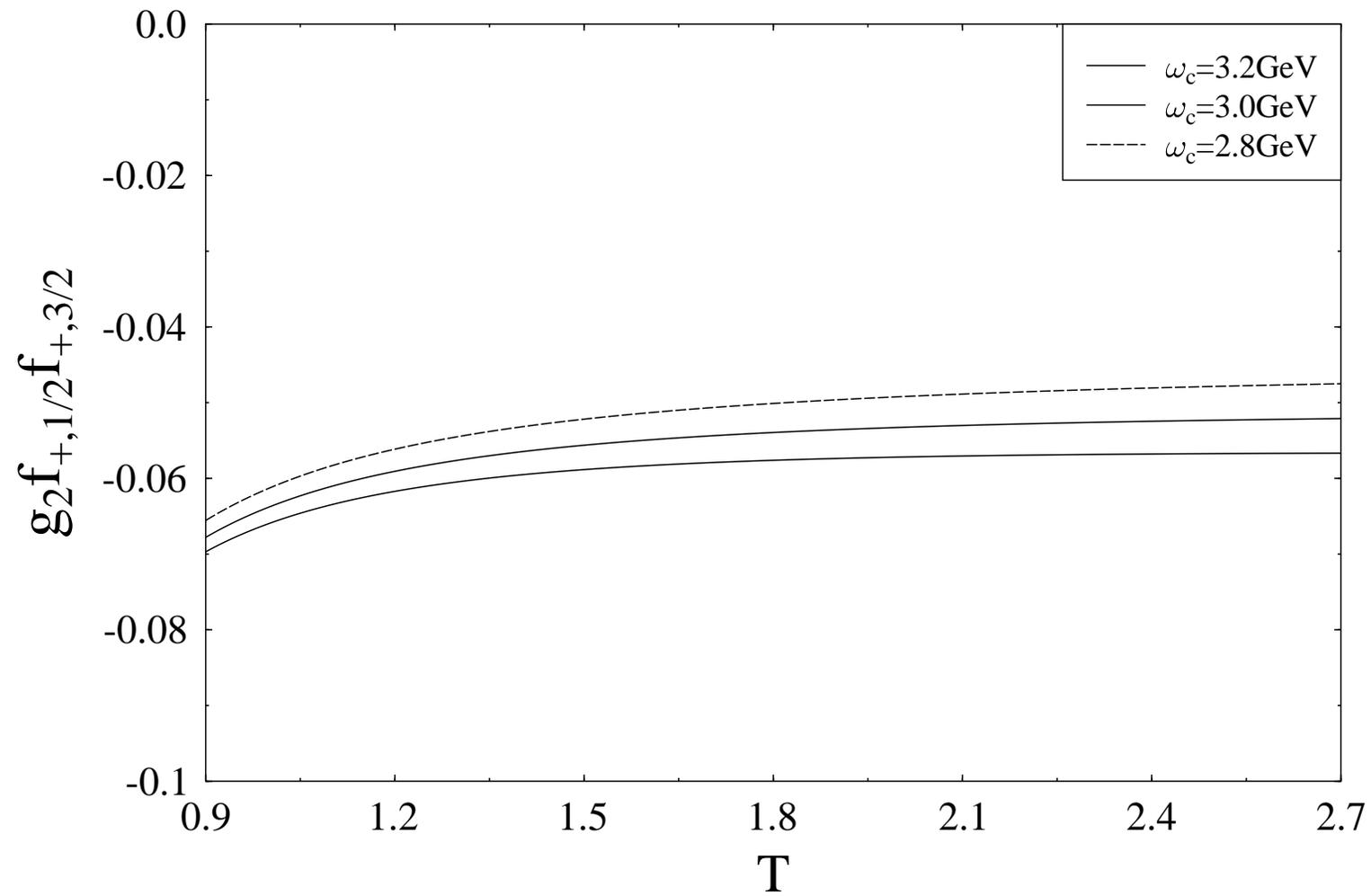
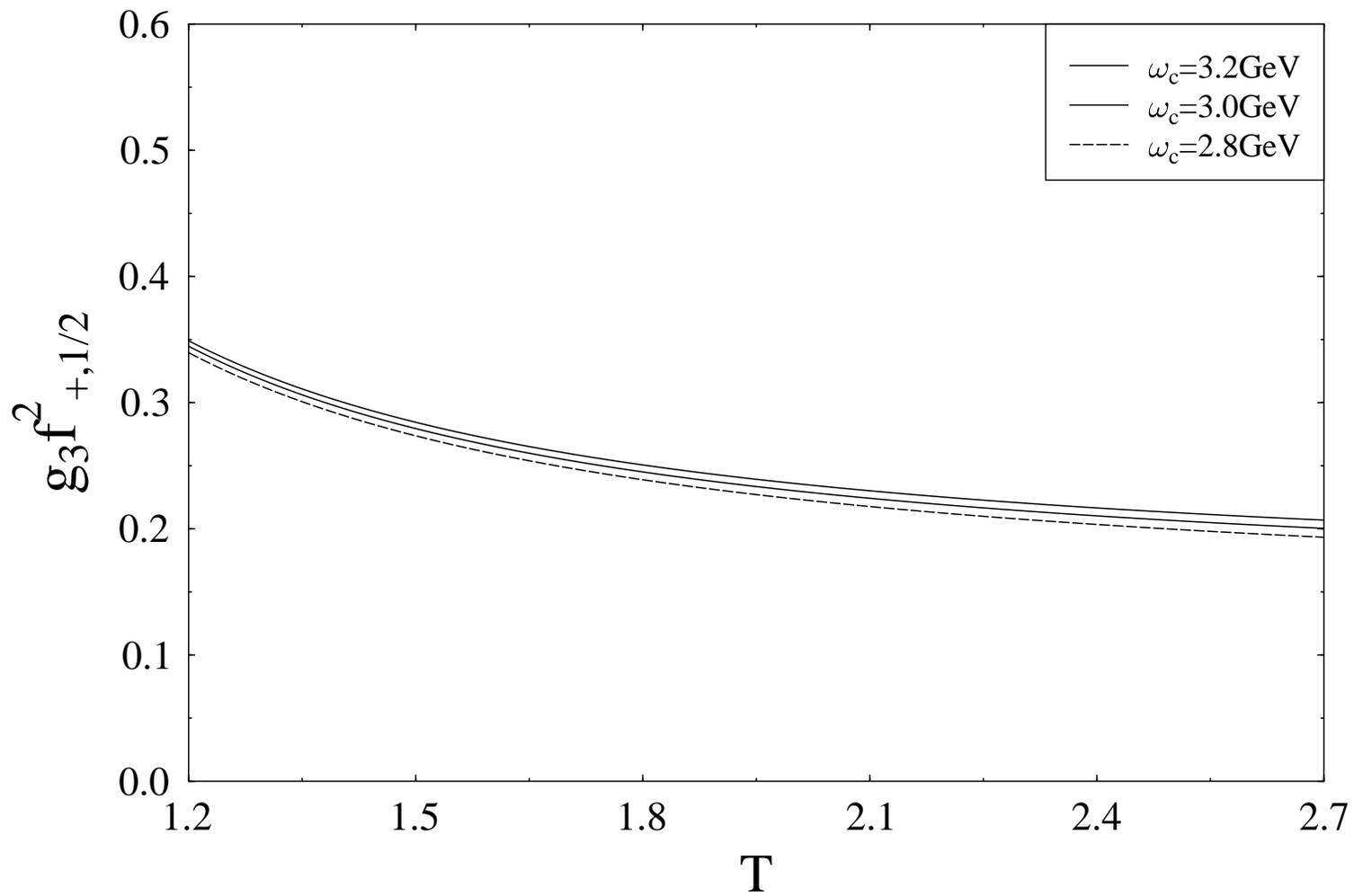
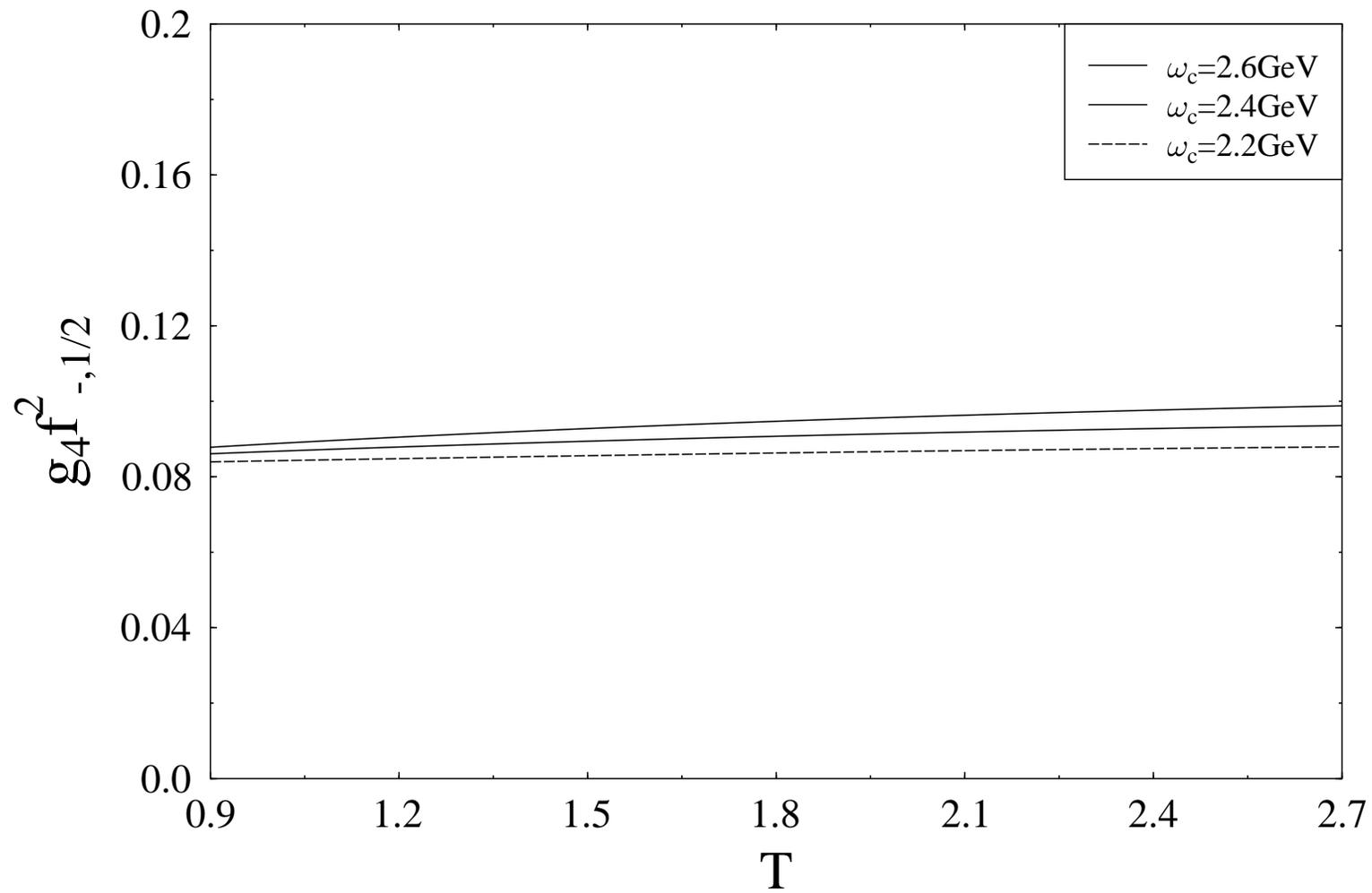
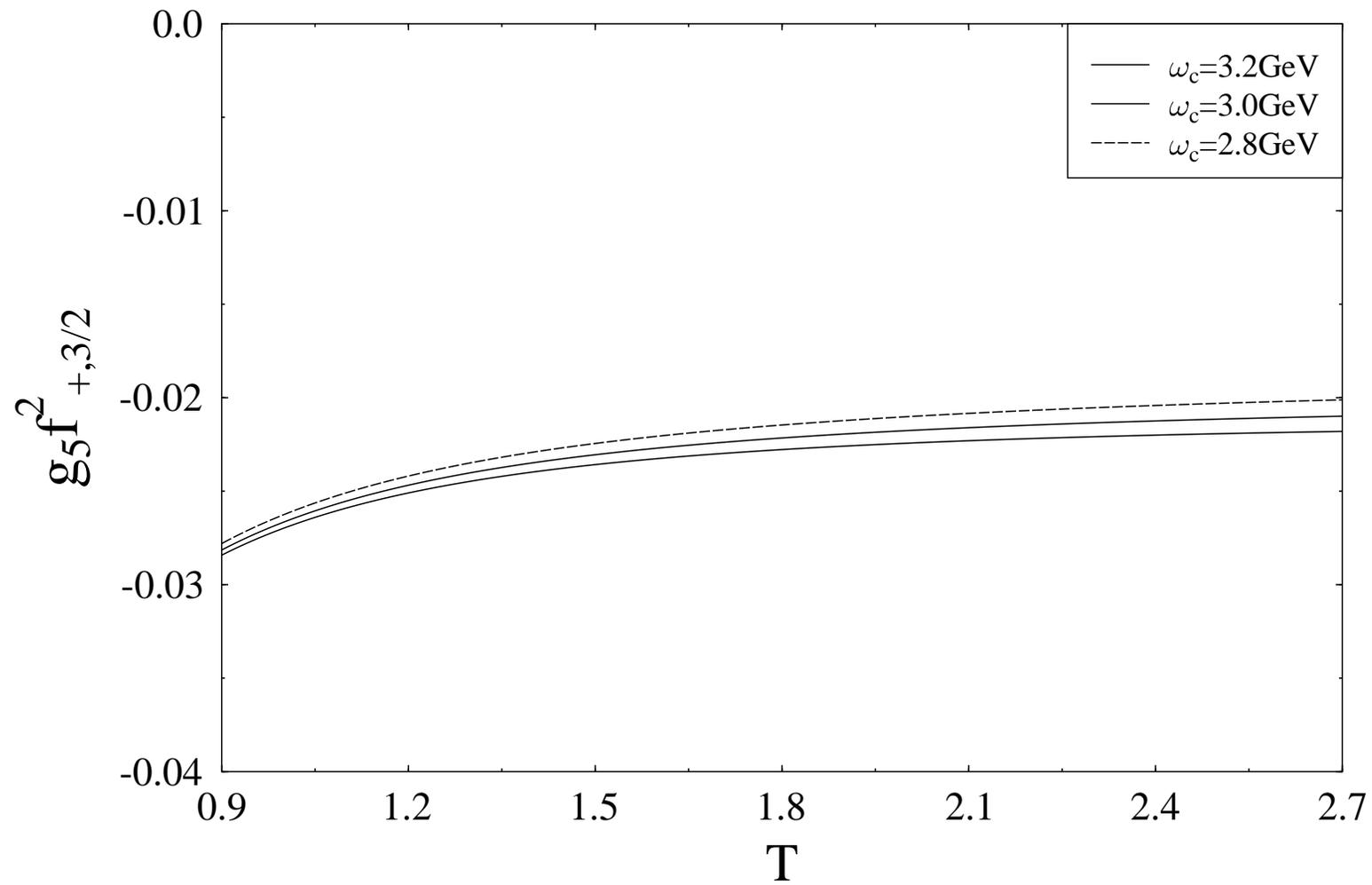
\end{document}